\documentclass[twocolumn]{aastex63}
\usepackage{amsmath}
\usepackage{xcolor}
\usepackage{physics}
\newcommand{\dmhalo}{DM$_{\text{halo}}$}
\newcommand{\dmdisk}{DM$_{\text{disk}}$}
\newcommand{\dmigm}{DM$_{\text{cosmic}}$}
\newcommand{\dmgal}{DM$_{\text{Gal}}$}
\newcommand{\dmhost}{DM$_{\text{host}}$}
\newcommand{\halomass}{$M_{\text{halo}}$}
\newcommand{\dmunits}{pc\,cm$^{-3}$}

\newcommand{\dmmeightyone}{87.8}

\newcommand{\anedit}[1]{#1}

\shorttitle{FRB Halo Plasma Constraints}
\shortauthors{Cook et al.}
\graphicspath{{./}{figures/}}

\begin{document}

\title{An FRB Sent Me a DM: Constraining the Electron Column of the Milky Way Halo with Fast Radio Burst Dispersion Measures from CHIME/FRB}

\correspondingauthor{Amanda M. Cook}
\email{cook@astro.utoronto.ca}

\author[0000-0001-6422-8125]{Amanda M. Cook}
\affiliation{David A. Dunlap Institute Department of Astronomy \& Astrophysics, University of Toronto, 50 St. George Street, Toronto, Ontario, Canada M5S 3H4}
\affiliation{Dunlap Institute for Astronomy \& Astrophysics, University of Toronto, 50 St. George Street, Toronto, Ontario, Canada M5S 3H4}

\author[0000-0002-3615-3514]{Mohit Bhardwaj}
\affiliation{McGill Space Institute, McGill University, 3550 rue University, Montr\'eal, QC H3A 2A7, Canada}
\affiliation{Department of Physics, McGill University, 3600 rue University, Montr\'eal, QC H3A 2T8, Canada}
\affiliation{Department of Physics, Carnegie Mellon University, 5000 Forbes Avenue, Pittsburgh, 15213, PA, USA}

\author[0000-0002-3382-9558]{B. M. Gaensler}
\affiliation{Dunlap Institute for Astronomy \& Astrophysics, University of Toronto, 50 St. George Street, Toronto, Ontario, Canada M5S 3H4}
\affiliation{David A. Dunlap Institute Department of Astronomy \& Astrophysics, University of Toronto, 50 St. George Street, Toronto, Ontario, Canada M5S 3H4}

\author[0000-0002-7374-7119]{Paul Scholz}
\affiliation{Dunlap Institute for Astronomy \& Astrophysics, University of Toronto, 50 St. George Street, Toronto, Ontario, Canada M5S 3H4}

\author[0000-0003-3734-8177]{Gwendolyn M. Eadie}
\affiliation{David A. Dunlap Institute Department of Astronomy \& Astrophysics, University of Toronto, 50 St. George Street, Toronto, Ontario, Canada M5S 3H4}
\affiliation{Department of Statistical Science, University of Toronto, Ontario Power Building, 700 University Avenue, 9th Floor, Toronto, ON, Canada M5G 1Z5}

\author[0000-0001-7301-5666]{Alex S. Hill}
\affiliation{Dominion Radio Astrophysical Observatory, Herzberg Research Centre for Astronomy and Astrophysics, National Research Council Canada, PO Box 248, Penticton, BC V2A 6J9, Canada}
\affiliation{Department of Physics and Astronomy, University of British Columbia, 6224 Agricultural Road, Vancouver, BC V6T 1Z1 Canada}

\author[0000-0001-9345-0307]{Victoria M. Kaspi}
\affiliation{McGill Space Institute, McGill University, 3550 rue University, Montr\'eal, QC H3A 2A7, Canada}
\affiliation{Department of Physics, McGill University, 3600 rue University, Montr\'eal, QC H3A 2T8, Canada}

\author[0000-0002-4279-6946]{Kiyoshi W. Masui}
\affiliation{MIT Kavli Institute for Astrophysics and Space Research, Massachusetts Institute of Technology, 77 Massachusetts Ave, Cambridge, MA 02139, USA}
\affiliation{Department of Physics, Massachusetts Institute of Technology, 77 Massachusetts Ave, Cambridge, MA 02139, USA}

\author[0000-0002-8376-1563]{Alice P. Curtin}
\affiliation{McGill Space Institute, McGill University, 3550 rue University, Montr\'eal, QC H3A 2A7, Canada}
\affiliation{Department of Physics, McGill University, 3600 rue University, Montr\'eal, QC H3A 2T8, Canada}

\author[0000-0003-4098-5222]{Fengqiu Adam Dong}
\affiliation{Department of Physics and Astronomy, University of British Columbia, 6224 Agricultural Road, Vancouver, BC V6T 1Z1 Canada}

\author[0000-0001-8384-5049]{Emmanuel Fonseca}
\affiliation{Department of Physics and Astronomy, West Virginia University, PO Box 6315, Morgantown, WV 26506, USA}
\affiliation{Center for Gravitational Waves and Cosmology, West Virginia University, Chestnut Ridge Research Building, Morgantown, WV 26505, USA}

\author[0000-0002-3654-4662]{Antonio Herrera-Martin}
\affiliation{David A. Dunlap Institute Department of Astronomy \& Astrophysics, University of Toronto, 50 St. George Street, Toronto, Ontario, Canada M5S 3H4}
\affiliation{Department of Statistical Science, University of Toronto, Ontario Power Building, 700 University Avenue, 9th Floor, Toronto, ON, Canada M5G 1Z5}

\author[0000-0003-4810-7803]{Jane Kaczmarek}
\affiliation{Dominion Radio Astrophysical Observatory, Herzberg Research Centre for Astronomy and Astrophysics, National Research Council Canada, PO Box 248, Penticton, BC V2A 6J9, Canada}

\author[0000-0003-2116-3573]{Adam E. Lanman}
\affiliation{Department of Physics, McGill University, 3600 rue University, Montr\'eal, QC H3A 2T8, Canada}
\affiliation{McGill Space Institute, McGill University, 3550 rue University, Montr\'eal, QC H3A 2A7, Canada}

\author[0000-0002-5857-4264]{Mattias Lazda}
\affiliation{Department of Physics, McGill University, 3600 rue University, Montr\'eal, QC H3A 2T8, Canada}

\author[0000-0002-4209-7408]{Calvin Leung}
\affiliation{MIT Kavli Institute for Astrophysics and Space Research, Massachusetts Institute of Technology, 77 Massachusetts Ave, Cambridge, MA 02139, USA}
\affiliation{Department of Physics, Massachusetts Institute of Technology, 77 Massachusetts Ave, Cambridge, MA 02139, USA}

\author[0000-0001-8845-1225]{Bradley W. Meyers}
\affiliation{Department of Physics and Astronomy, University of British Columbia, 6224 Agricultural Road, Vancouver, BC V6T 1Z1 Canada}
\affiliation{International Centre for Radio Astronomy Research (ICRAR), Curtin University, Bentley WA 6102 Australia}

\author[0000-0002-2551-7554]{Daniele Michilli}
\affiliation{MIT Kavli Institute for Astrophysics and Space Research, Massachusetts Institute of Technology, 77 Massachusetts Ave, Cambridge, MA 02139, USA}
\affiliation{Department of Physics, Massachusetts Institute of Technology, 77 Massachusetts Ave, Cambridge, MA 02139, USA}

\author[0000-0002-8897-1973]{Ayush Pandhi}
\affiliation{David A. Dunlap Institute Department of Astronomy \& Astrophysics, University of Toronto, 50 St. George Street, Toronto, Ontario, Canada M5S 3H4}
\affiliation{Dunlap Institute for Astronomy \& Astrophysics, University of Toronto, 50 St. George Street, Toronto, Ontario, Canada M5S 3H4}

\author[0000-0002-8912-0732]{Aaron B. Pearlman}
\affiliation{Department of Physics, McGill University, 3600 rue University, Montr\'eal, QC H3A 2T8, Canada}
\affiliation{McGill Space Institute, McGill University, 3550 rue University, Montr\'eal, QC H3A 2A7, Canada}

\author[0000-0002-4795-697X]{Ziggy Pleunis}
\affiliation{Dunlap Institute for Astronomy \& Astrophysics, University of Toronto, 50 St. George Street, Toronto, Ontario, Canada M5S 3H4}

\author[0000-0001-5799-9714]{Scott Ransom}
\affiliation{National Radio Astronomy Observatory, 520 Edgemont Rd, Charlottesville, VA 22903, USA}

\author[0000-0003-1842-6096]{Mubdi Rahman}
\affiliation{Sidrat Research, PO Box 73527 RPO Wychwood, Toronto, ON M6C 4A7, Canada}
\author[0000-0003-3154-3676]{Ketan R. Sand}
\affiliation{Department of Physics, McGill University, 3600 rue University, Montr\'eal, QC H3A 2T8, Canada}
\affiliation{McGill Space Institute, McGill University, 3550 rue University, Montr\'eal, QC H3A 2A7, Canada}

\author[0000-0002-6823-2073]{Kaitlyn Shin}
\affiliation{MIT Kavli Institute for Astrophysics and Space Research, Massachusetts Institute of Technology, 77 Massachusetts Ave, Cambridge, MA 02139, USA}
\affiliation{Department of Physics, Massachusetts Institute of Technology, 77 Massachusetts Ave, Cambridge, MA 02139, USA}

\author[0000-0002-2088-3125]{Kendrick Smith}
\affiliation{Perimeter Institute for Theoretical Physics, 31 Caroline Street N, Waterloo, ON N25 2YL, Canada}

\author[0000-0001-9784-8670]{Ingrid Stairs}
\affiliation{Department of Physics and Astronomy, University of British Columbia, 6224 Agricultural Road, Vancouver, BC V6T 1Z1 Canada}
\author[0000-0002-9761-4353]{David C. Stenning}
\affiliation{Department of Statistics \& Actuarial Science, Simon Fraser University, Burnaby, BC, Canada}
\begin{abstract}
The CHIME/FRB project has detected hundreds of fast radio bursts (FRBs), providing an unparalleled population to probe statistically the foreground media that they illuminate. One such foreground medium is the ionized halo of the Milky Way (MW).
We estimate the total Galactic electron column density from FRB dispersion measures (DMs) as a function of Galactic latitude using four different estimators, including ones that assume spherical symmetry of the ionized MW halo and ones that imply more latitudinal-variation in density. Our observation-based constraints of the total Galactic DM contribution for $|b|\geq30^\circ$, depending on the Galactic latitude and selected model, span 87.8$\,-\,141$\,pc\,cm$^{-3}$. This constraint implies upper limits on the MW halo DM contribution that range over $52-111$\,pc\,cm$^{-3}$.
 We discuss the viability of various gas density profiles for the MW halo that have been used to estimate the halo's contribution to DMs of extragalactic sources. Several models overestimate the DM contribution, especially when assuming higher halo gas masses ($\sim 3.5 \times 10^{12} M_\odot$). Some halo models predict a higher MW halo DM contribution than can be supported by our observations unless the effect of feedback is increased within them, highlighting the impact of feedback processes in galaxy formation. 


\end{abstract}
\section{Introduction}
\label{sec:intro}


Our Galactic halo connects the baryon-rich intergalactic medium (IGM) to the disk of the Milky Way (MW). Gas from the halo is a combination of new and recycled material, is a consequence of galactic feedback processes, and represents a galaxy's future star formation fuel. The MW halo contains both neutral and ionized gas, although it is dominated in mass by the latter component, which extends to hundreds of kiloparsecs \citep{1991IAUS..144...67R, 2012ARA&A..50..491P}. 

Meaningful theoretical predictions of the composition and size of the MW halo come from our knowledge of cosmology and galaxy formation theory (for a review, see \citealt{2012ARA&A..50..491P}). In this sense, the total mass and extent of the halo can be used to check our understanding of these topics. Unfortunately, due to the diffuse and hot nature of the ionized halo gas and our position within the MW, the MW halo gas cannot be imaged directly. 

Existing indirect constraints for the total amount of hot halo gas have been placed using observations of the $\sim 0.1-1$ keV diffuse soft X-ray background, which find typical emission measures of $(1.4-3.0) \times 10^{-3}$\,cm$^{-6}$\, pc \citep{ 2009ApJ...707..644G, 2009PASJ...61..805Y, 2010ApJ...723..935H,2013ApJ...773...92H}. Indirect constraints for the total amount of plasma have also been placed using absorption lines of oxygen ions in X-ray and far-ultraviolet spectroscopy of active galactic nuclei, \anedit{however, there are considerably fewer useful sight lines \citep{2015ApJS..217...21F, 2012ApJ...756L...8G, 2012AIPC.1427..342S}. 
 Most of the hot gas detected in X-ray emission is thought to be within a few kiloparsecs of the MW disk \citep{2006ApJ...644..174F,2007ApJ...658.1088Y}, although evidence exists for extended hot halo gas with density on the order of $10^{-5} - 10^{-4}\,\text{cm}^{-2}$ at distances of 50$-$100 kpc \citep{2003ApJS..146..165S,2006ApJ...653.1210S,2009ApJ...696..385G}. More accurate estimates of the total mass of the ionized medium in the halo require a more precise knowledge of the physical properties of this extended ionized gas. Evidence is emerging suggesting more structure within the MW halo gas, although most models previously had assumed spherical symmetry. \cite{2020ApJ...888..105Y} and \cite{2022PASJ...74.1396U} find evidence for a disk-like component to the MW halo gas. The gas closest to us (within 50 Mpc) suggests that the hot halo gas cannot be the host of all of the missing baryons \citep{2015JATIS...1d5003B, 2018ApJ...862....3B}. However, \cite{2017ApJ...835...52F} show that if the gas density beyond 50\,kpc were to flatten, the hot halo gas could account for the missing baryons. }

Another constraint on the mass and extent of the plasma in the MW halo comes from radio observations of pulsars in the Large and Small Magellanic Clouds  \citep[LMC and SMC; e.g.,][]{2013MNRAS.433..138R}. The LMC \& SMC are located $\sim$ 50 and 60 kpc away, respectively  \citep{2019Natur.567..200P, 2020ApJ...904...13G}, but this distance is only a small fraction of the virial radius of the MW (using the definition of \citealt{1998ApJ...495...80B}, current estimates of the latter are \anedit{typically between 180 and 250\,kpc \citep{2015ApJS..216...29B,2020MNRAS.494.4291C,2022ApJ...925....1S}).}

The key to the LMC and SMC pulsar-based halo constraint is the significant dispersive effect of ionized gas on radio waves. 
Precise measurements of arrival times at the top (high frequencies, $\nu_1$) versus bottom (low frequencies, $\nu_2$) of a radio survey's observing band or of the detectable emission for short bursts allow for a quantification of the dispersive delay known as the dispersion measure (DM). This effect is mainly due to the electrons along the line of sight and thus approximately (i.e., good to within one part per thousand, see \citeauthor{shridm} \citeyear{shridm} for an in-depth discussion) proportional to the column density of free electrons,

\begin{equation}
    \text{DM} = \int_0^L n_e \ \dd l 
\end{equation}

\noindent where $L$ is the distance to the source, in this case the pulsars in the LMC or SMC, and $n_e$ is the free electron number density. DM is determined from observations via the relationship

\begin{equation}
    \Delta t = \frac{(e^{-})^2}{2 \pi m_e c } \left( \frac{1}{\nu_2^2} - \frac{1}{\nu_1^2} \right) \text{DM}
\end{equation}

\noindent where $\Delta t$ is the wave arrival time delay between $\nu_1$ and $\nu_2$\footnote{The first term of constants is set to be exactly $1/2.41\times10^{-4}$ within CHIME/FRB's pipeline \citep{overview}.}. DM is a direct probe of the intervening plasma between observers and radio transients. The radio pulsars within the SMC and LMC set a lower bound on the Galactic DM contribution at their respective distances of $70\pm 3$ and $45\pm1$ \dmunits \ \citep{2006ApJ...649..235M}.

DM is also useful for constraining the plasma within the Galactic disk. One can characterize this medium using pulsars with independent distance measurements, typically through annual parallax, which enables the modelling of the scale height and midplane density of the warm ionized medium (WIM) disk \citep{ne2001, 2008PASA...25..184G, 2009ApJ...702.1472S, 10.1111/j.1365-2966.2012.21869.x, ymw16, ocker}. 

Galactic plasma models NE2001 \citep{ne2001} and the more recent YMW16 \citep{ymw16} include more components than scale height, filling factor, and vertical electron column, in contrast to the other models listed above, but NE2001 and YMW16 are also based on DM measurements of pulsars with independent distance measurements. Both models include components for the thin and thick disk, spiral arms, and local structures like the Local Bubble and the Gum Nebula. NE2001 model parameters were fit using data from 112 pulsar distances and 269 scattering measurements. YMW16 used 189 independent pulsar distances and an updated estimate of the WIM disk scale height\footnote{\cite{ymw16} argued that scattering measures are generally dominated by a few foreground structures along the line of sight to a pulsar and not large scale structure of the Galaxy, so they did not include these measurements in their model. Notably, neither model includes a component for the MW halo.}. Both models have been shown to fail in predicting the DMs of certain populations like high latitude pulsars, pulsars in \ion{H}{2} regions, and several relatively-local pulsars \citep{chatterjee_VLBA}. \cite{2021arXiv210615816P} give a comprehensive review and comparison of these two models. 

Unfortunately, there are very few known pulsars available to probe significant fractions of the MW halo. One expects to find the highest density of canonical pulsars, remnants of short-lived massive stars, within the disk, and hence historical pulsar surveys most commonly target this area. Typical pulsar emission is also too faint to readily observe at great distances like the edge of the Galaxy or beyond.

Fast radio burst (FRB) DMs are a new way to constrain the total mass and extent of the halo. The class-defining observation of an FRB \citep{lorimer}, FRB 20010724A caught the attention of astronomers in part because the burst had a DM higher than could be contributed by our Galaxy along that line of sight according to Galactic electron density models like NE2001.


Most observed FRBs have DMs many times what these models estimate can be expected from our Galaxy based on the above models or on the measured scale height and average vertical electron column of the MW. In all published instances of precise FRB localizations to date, the FRB is spatially coincident with a galaxy (for a review of host galaxy associations, see \citealt{2020ApJ...903..152H}). These associations confirm their extragalactic nature as the chance coincidence of finding a galaxy that is physically unrelated to the source in their small localization regions is negligible. FRBs with DMs substantially larger than that maximally predicted by Galactic density models in their line of sight can be assumed to be extragalactic\footnote{There have been detections of FRB-like events from within the MW from a known Galactic source, namely, magnetar SGR 1935+2154 \citep{sgrchime, sgrchris, 2022ATel15681....1D}. The DM of this burst was consistent with being Galactic according to NE2001 and YMW16.}.

We can define the measured DM of an extragalactic FRB as the sum of the following four components:
\begin{equation}
    \text{DM} = \text{DM}_{\text{disk}} + \text{DM}_{\text{halo}} + \text{DM}_{\text{cosmic}} + \frac{\text{DM}_{\text{host}}}{1+z} \label{eqn:dmcont}
\end{equation}

 \noindent where the terms refer to the DM contributions from electrons in the MW disk, the MW halo, the \anedit{cosmic web}, and the FRB host galaxy. 
 The first two terms, \dmdisk\ and \dmhalo \ when summed are denoted \dmgal \ as they comprise the contribution from the MW in a given line of sight, \anedit{i.e.,}
\begin{equation}
    \text{\textbf{DM}}_{\text{\textbf{Gal}}} = \text{\textbf{DM}}_{\text{\textbf{disk}}} + \text{\textbf{DM}}_{\text{\textbf{halo}}}.
\end{equation}
  \dmhost \ likely includes contributions from the halo and disk of the host galaxy, and potentially includes a local component around the source of the burst. \anedit{\dmigm \ includes contributions from the IGM,} contributions from ionized gas in our Local Group \citep{prochaskaandzhang}, and could include intervening galaxies or galaxy halos along the line of sight to the FRB \citep{2019Sci...366..231P}.


The MW halo contribution was first constrained using a population of FRBs by \cite{platts}. After subtracting the \dmdisk\ estimated by NE2001 from the total measured DM of each FRB, the authors model the excess DM distributions using asymmetric kernel density estimation and set conservative limits $-2 < $\,\dmhalo \,$<123$\,\dmunits. The authors concluded by emphasizing that they expect a larger sample of FRBs will tighten these constraints. 

In this paper we derive observation-based upper limits of \dmhalo \ as a function of Galactic latitude from the most extensive sample of FRBs to date. In Section \ref{sec:data}, we outline the extragalactic source sample from the FRB backend of Canadian Hydrogen Intensity Mapping Experiment (CHIME), which allows us to make direct upper limits of the column density of ionized halo gas without relying on models for \dmhalo, \dmigm, and \dmhost, each of which remain loosely constrained on a population scale. 
In Section \ref{sec:results}, we compare this extragalactic sample with information from Galactic pulsar DMs and show that there is a distinct gap between the extragalactic and Galactic populations. 
Then, in Section \ref{sec:analysis}, we derive estimates of \dmgal \ as a function of Galactic latitude which, when combined with estimates of \dmdisk, also describe the structure of \dmhalo \ as a function of Galactic latitude.  
We discuss the biases in our data collection in Section \ref{sec:biases} \anedit{and how these biases could produce the lack of radio pulse detections between our Galactic and extragalactic populations}. The uncertainties of our derived models are discussed in Section \ref{sec:errors}. 

\section{Fast Radio Burst Sample}
\label{sec:data} 
Our extragalactic FRB sample comes from CHIME/FRB. CHIME is a radio telescope operating over 400--800 MHz \citep{2022ApJS..261...29C}. CHIME is a transit telescope with no moving parts; it observes the sky above it as the Earth rotates. CHIME is located at the Dominion Radio Astrophysical Observatory near Penticton, British Columbia, Canada. The CHIME telescope is comprised of four 20m $\times$ 100m, North/South oriented, semi-cylindrical parabolic reflectors, each of which has 256 dual-polarization feeds, giving the entire instrument a more than 200-square-degree field of view. CHIME's FX correlator forms 1024 beams over this large field of view, and and the FRB backend searches the beams for radio pulses with durations of $\sim 1$ to hundreds of milliseconds, such as pulsars and FRBs \citep{overview}.
For this study, we selected all 93 sources detected by CHIME/FRB through February 2021 that satisfied our selection criteria; namely, having a low measured DM ($<250$ \dmunits) and high Galactic latitude ($|b|>30^\circ$). 34 of these FRBs are reported in the first CHIME/FRB catalog \citep{catalog}. We inspected the events for any evidence that they were detected away from the meridian of the telescope (i.e., in a sidelobe), as this can result in a given burst's reported position being inaccurate due to imperfect modeling of inherent sidelobe-beam structure. None of the bursts in our sample show evidence of being sidelobe events, but especially with the lower S/N bursts we cannot completely rule out this possibility with intensity data alone. 

We define high latitude as those FRBs with measured absolute Galactic latitude ($|b|$) greater than 30$^{\circ}$. We made this selection to avoid contamination in measured DM by \ion{H}{2} regions and other small scale local structures. At these latitudes the maximal \dmgal \ predicted by Galactic free electron density models YMW16 and NE2001 show significantly less scatter in Galactic longitude, a dimension we collapse over in this study. 
\begin{figure*}
\plotone{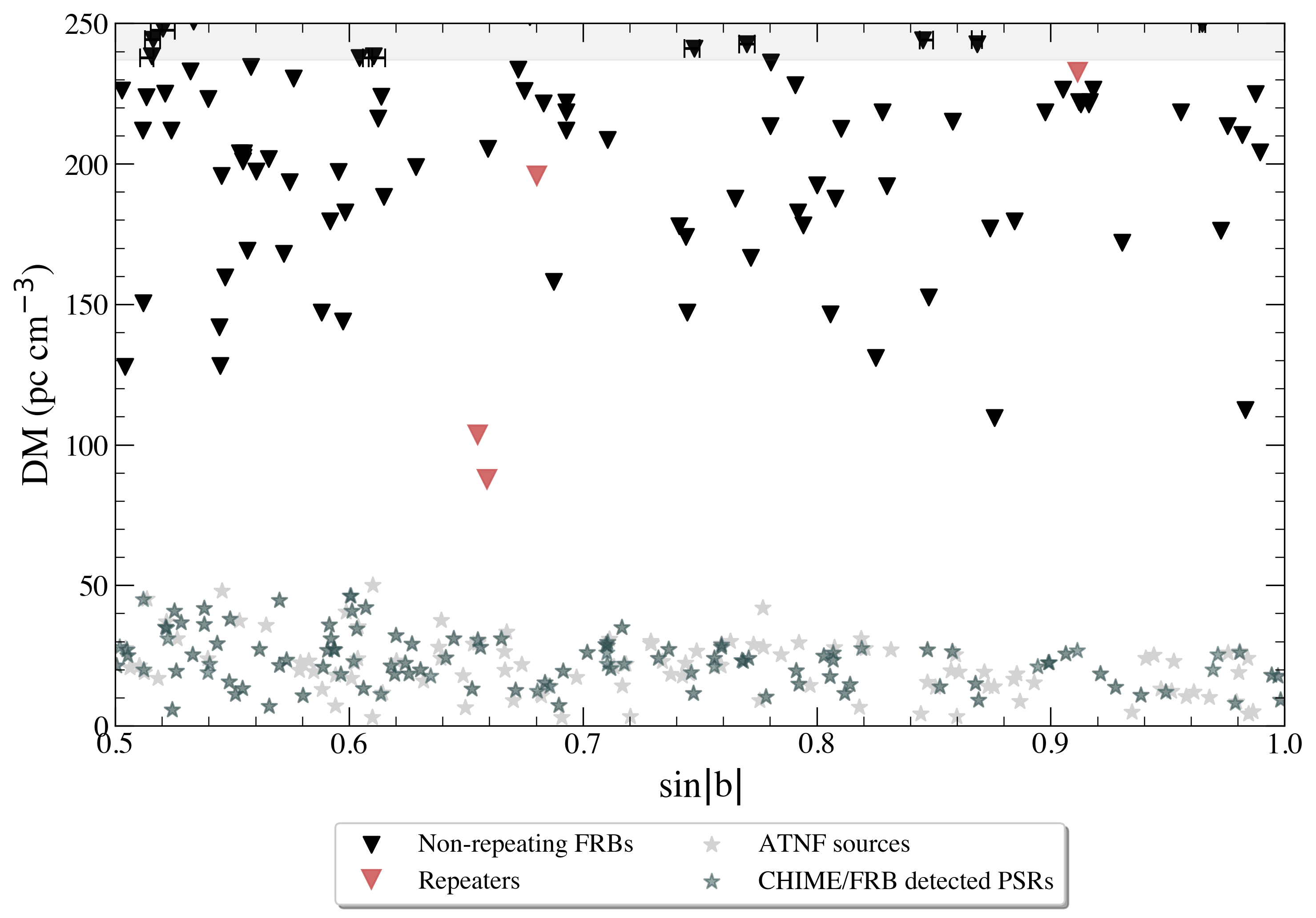}
\caption{
Total measured DM as a function of $\sin|b|$ for $|b| \geq 30^\circ$ for FRBs detected by CHIME/FRB with DM less than 250\,\dmunits \ through February 2021. Non-repeating FRBs are represented with black triangles and repeating FRB sources represented with red triangles. Galactic sources, namely pulsars from the ATNF Pulsar Catalogue (light gray) \citep{atnf} and all Galactic sources detected by CHIME/FRB's realtime pipeline (dark gray) are shown \citep{ 2021ApJ...922...43G}.  We do not plot, however, sources from lines of sight with very high emission measure as measured by Planck \citep{planckfreefree} to avoid higher-than-representative DMs due to contamination by \ion{H}{2} regions and other small scale, local structure. Similarly, sources with declination $<-11^{\circ}$ are not plotted as they are outside of CHIME/FRBs field-of-view, such that longitudinal variation is comparable between the Galactic and extragalactic samples. Representative positional errors are shown for sources in the top gray band.  The DM errors of the FRBs are much smaller than the markers so we do not plot them. A clear gap in DM is visible between the triangles and stars.}
\label{fig:nolines}
\end{figure*}
CHIME/FRB's pipeline imposes another selection criterion on our sample. The pipeline only saves intensity data from bursts with measured DMs greater than at least one of YMW16 or NE2001's maximal \dmdisk \ estimate in the burst's apparent line of sight. The impact of this pipeline-imposed selection criterion is discussed further in Section \ref{sec:biases}. Our low DM sample at these latitudes are FRBs with DM $< 250$\,\dmunits \ and are chosen as they are the most constraining on \dmhalo. Most MW halo models typically translate into \dmhalo \ predictions less than 100 \dmunits, and at Galactic latitudes greater than $30^{\circ}$, \dmdisk \ is predicted to be less than 70\,\dmunits \ according to NE2001, YMW16, and \cite{ocker}. Thus, we choose to consider only FRBs with measured DM less than $250$\, \dmunits \ to conservatively explore the range within which models predict \dmgal. 

Our selected sample includes four repeating sources, one of which, FRB 20200120E, is associated with spiral galaxy M81 and at 3.6 Mpc away, is the closest known extragalactic FRB source \citep{m81r, m81rloc}. FRB 20200120E, which we will denote M81R for brevity, is a particularly interesting source for this study, not only because it likely has a low \dmigm \ contribution (\citealt{m81rloc} estimate this contribution to be on the order of 1 \dmunits), but also because it is located in a globular cluster on the outskirts of M81 (the globular cluster's offset from the center of M81, measured in projection, is approximately 20 kpc). This circumstance means that we expect a negligible DM contribution from the disk of M81. Additionally, we do not expect that a globular cluster would contribute significant amounts of internal dispersion \citep{2001ApJ...557L.105F}. 




\section{Galactic and Extragalactic Comparisons}
\label{sec:results}
The extragalactic FRBs with the lowest DMs provide the most constraining upper limits on \dmgal. Figure \ref{fig:nolines} shows the DMs of all high-latitude and low-DM FRB candidates from CHIME/FRB (triangles) as a function of $\sin|b|$. Repeating FRB sources are shown as red triangles, indicating the best measured latitude and DM considering all published bursts. The FRB with the smallest DM in our current sample is M81R with DM\,=\,\dmmeightyone\,\dmunits. 

We plot all Galactic pulsars in DM vs Galactic latitude from the Australia Telescope National Facility (ATNF) pulsar catalog\footnote{Version: 1.64, Accessed: 23/03/2021,  \url{http://www.atnf.csiro.au/research/pulsar/psrcat}} \citep{atnf} (light gray stars) and indicate the sources from this sample that have been detected by the realtime CHIME/FRB pipeline (dark teal stars). Additionally, if pre-publication pulsars or RRATs from the pulsar survey scraper\footnote{\url{https://pulsar.cgca-hub.org/}} have been detected by CHIME/FRB's realtime pipeline through February 2021, we also include them in this plot (dark teal stars). This addition includes new Galactic sources seen by CHIME\footnote{see \url{https://www.chime-frb.ca/galactic} \anedit{for the most up-to-date catalog}} \citep{2021ApJ...922...43G, GWG2}. We exclude ATNF and pulsar survey scraper sources that were detected in lines of sight with emission measures above the 95$^{\text{th}}$ percentile of the sky as measured by the Planck 2015 astrophysical component separation analysis \citep{planckfreefree}. This exclusion is enacted to avoid higher-than-representative DMs due to contamination by \ion{H}{2} regions\footnote{Most \ion{H}{2} are located at low absolute Galactic latitudes, but there are a few which have been observed at Galactic latitudes relevant to our study \citep{2003A&A...397..213P}.} and other small scale, local structure. This cut affects about 30\% of pulsars across the sky, but does not remove any pulsars from the Galactic latitudes and declinations we consider. The pulsar declination criterion removes sources that are outside of CHIME's sky coverage (i.e., those with declinations $<-11^\circ$). The pulsars and RRATs all sit below a DM of $\approx 50$ \dmunits, and the highest pulsar or RRAT DMs are largely found at lower $\sin|b|$. 

There is a distinct gap in DMs at around 50--\dmmeightyone\,\dmunits \ (the exact values depend on the latitude considered, but the gap is not narrower than this in any direction). As discussed more in Section \ref{sec:biases}, for $|b|>30^{\circ}$ this gap separates known or suspected Galactic sources and the extragalactic FRBs. 

\section{Analysis}
\label{sec:analysis}

\subsection{\anedit{Basic Methodology}}

We seek to describe the constraints placed on \dmgal \ using FRBs as a function of Galactic latitude. \anedit{Reasonable possibilities for models of \dmgal\ include those which assume a purely spherical ionized MW halo and models which imply more latitudinal-variation in the density of the ionized MW halo. We will fit a model which assumes \dmgal\ is constant, a model that assumes \dmhalo\ is constant, and models for \dmgal\ which take the form of third-order polynomials but still bound the DMs of the FRBs from below. Since measured extragalactic FRB DMs must include the contribution from \dmgal\ along their line of sight, and each of these models bound the FRB DMs from below, the models represent the upper limits of \dmgal\ derived when \dmigm = \dmhost$= 0$. We then turn the upper limit models of \dmgal\ at each latitude into upper limits of \dmhalo\ by subtracting \dmdisk\ component as a function of Galactic latitude ($b$) found by \cite{ocker},}
\begin{equation}
    \text{DM}_{\mathrm{disk}} = \frac{23.5 \pm 2.5}{\sin|b|} \ \text{pc\,cm}^{-3}. \label{eqn:ocker}
\end{equation}
\subsection{\anedit{Galactic DM Estimates}}
\begin{figure*}
\plotone{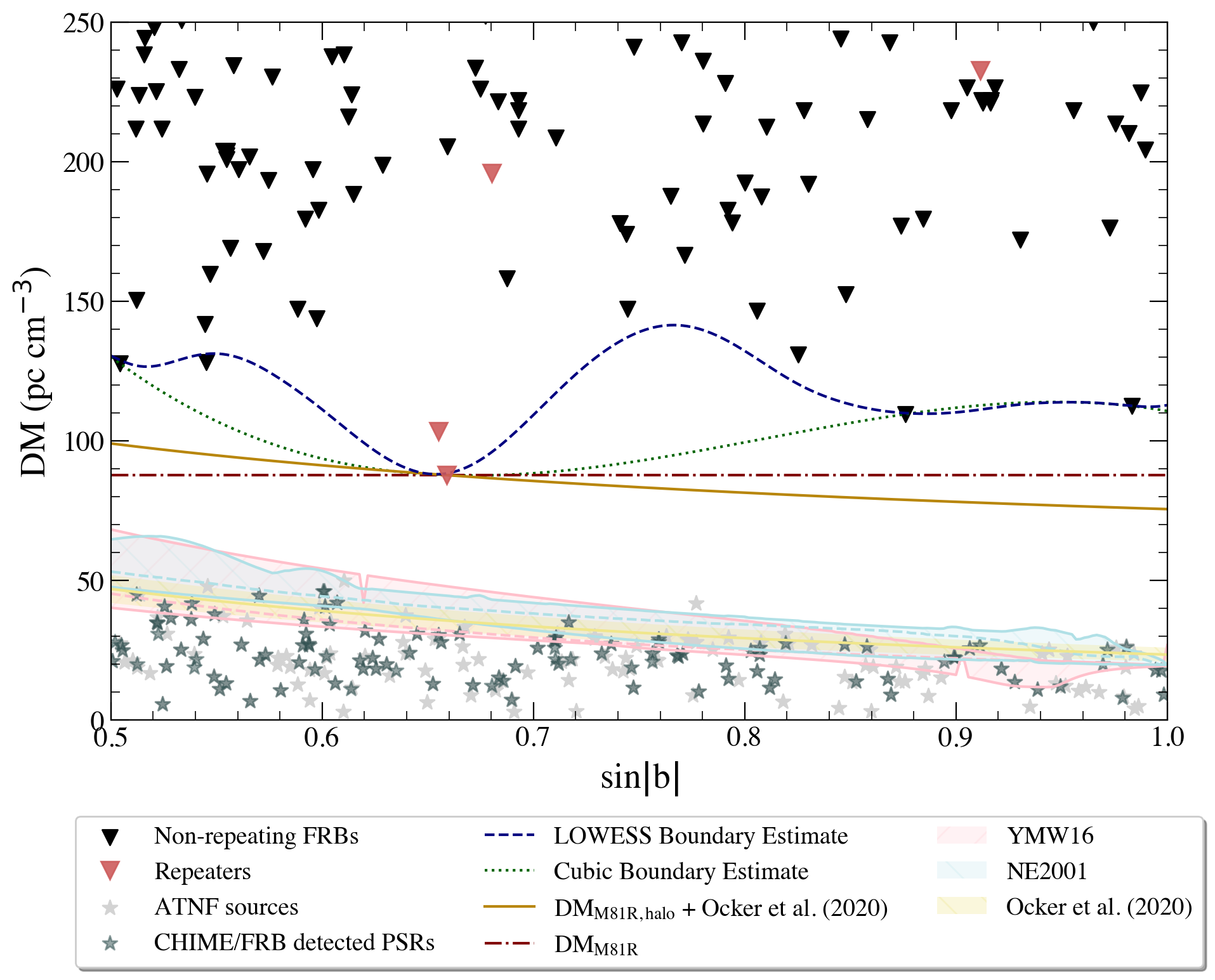}
\caption{As for Figure \ref{fig:nolines}, but four simple boundary models of \dmgal \ are shown, which display the most conservative estimates supported by CHIME/FRBs extragalactic DM sample, using different fitting methods and polynomial degrees (see Section \ref{sec:analysis} for details). Additionally, the total expected Galactic contribution to the DM from the two Galactic free electron density models, NE2001 and YMW16, are plotted in blue and pink respectively, where the shaded regions bounded by solid lines represent the range in values for lines of sight which vary with Galactic longitude \citep{ne2001, ymw16}. The pink, yellow, and blue dotted lines show the median value of the YMW16, \cite{ocker}, and NE2001 respectively at each Galactic latitude. 
The implied DM of the WIM disk component as a function of Galactic latitude found by \cite{ocker} is shown in yellow. 
}
\label{fig:highlat}
\end{figure*}

In Figure \ref{fig:highlat} we fit and plot four structure estimates that either strictly or roughly bound the DMs of FRBs from below, since the lowest extragalactic FRB DMs are the most constraining upper limits of the MW halo. The first estimate (red dot-dashed line in Figure \ref{fig:highlat}) assumes a constant value for the total Galactic contribution \dmgal \ across the sky. That is, \dmgal\,=\,\dmmeightyone\,\dmunits \ for all $|b| \in (30^{\circ},90^{\circ})$. This is the measured DM of FRB 20200120E, associated with spiral galaxy M81 \citep{m81r, m81rloc}. Below absolute latitudes of 30$^{\circ}$ this model is not supported, as many pulsars have been detected at DMs higher than \dmmeightyone\,\dmunits. 

The next model (solid yellow line in Figure \ref{fig:highlat}) assumes that the halo has a constant contribution at a given latitude. If \dmdisk \ is assumed to be the central value predicted by \cite{ocker} at the latitude of our lowest DM FRB, likely the most constraining single estimate of \dmhalo, our observations support a \dmhalo \ of no more than \dmmeightyone\,\dmunits $- (23.5/ \sin(0.64))$\,\dmunits $= 52$\, \dmunits. 

Written explicitly, 

\begin{equation}
\text{\dmgal}(b) = \left(\frac{23.5}{\sin|b|} + 52\right) \,\text{pc cm}^{-3}.
\end{equation}

We also fit a model for \dmgal \ using Locally Weighted Scatterplot Smoothing \citep[LOWESS;][]{lowess} to the local minimum of the measured FRB DMs (blue dashed line, Figure \ref{fig:highlat}). LOWESS is a method for smoothing a scatterplot in which the fitted value at a given point is the value of a polynomial fit to the data using weighted least squares. The weight is determined by how close the original value is to a local regression so that the weight is large if the proposed value is close to the data and small if not. At each point in $\sin|b|$ we bin all FRB DMs within 5$^{\circ}$ and select the minimum as the value for that latitude. Using these minima, we fit a LOWESS line with a polynomial degree of three and bandwidth of 0.55. A bandwidth of 0.55 means that 55\% of the data are considered when smoothing each point. The polynomial degree was chosen as both quadratic and quartic functions predicted behavior near the $|b|$ boundaries that were unphysical, and higher polynomial degrees didn't offer a better fit. The \dmgal \ predictions from this model fall between 88\,\dmunits \ at $\sin|b| = 0.65$ ($b = 40.8 ^{\circ}$) and 111\,\dmunits \ at $\sin|b| = 0.77$ ($50.1 ^{\circ}$). The LOWESS line demonstrates a lot of variation with Galactic latitude. This model is not intended to suggest a physical representation of structure in the halo. Rather, we wanted to demonstrate what conservative upper limits on \dmgal \ might be reasonable in lines-of-sight which do not have a particularly constraining FRB DM, using more constrained lines-of-sight nearby in Galactic latitude. These more constrained lines-of-sight are still relatively sparse at our sample size. This model is likely most appropriate if a conservative estimate is desired. 
\begin{deluxetable}{lc}
\label{tab:mod_coef}
\caption{Best fit parameters derived for the FRB DM cubic boundary estimate described in Equation \ref{eqn:cbe}.}
\tablehead{
\colhead{Coefficient} & \colhead{Value (pc\,cm$^{-3}$)}}
\startdata
$x_0$ & 1304 \\
$x_1$ & $-4747$ \\
$x_2$ &6044 \\
$x_3$ & $-2490$
\enddata
\end{deluxetable}

The final method we apply to model \dmgal \ as a function of Galactic latitude is polynomial boundary regression. The polynomial boundary regression assumes that the boundary of a given scatterplot can be described by a polynomial and optimizes that polynomial such that it envelopes the data and minimizes the area under its graph \citep{Hall_Park_Stern_1998}. We computed this estimate assuming a third degree polynomial using the CRAN package \textit{npbr}\footnote{\url{https://CRAN.R-project.org/package=npbr}} \citep{npbr}. 
For cubic polynomial with coefficients defined as 

\begin{equation}
\label{eqn:cbe}
   \text{DM}_{\text{Gal}} (b) = \sum_{i=0}^3 x_i\sin^i|b|,
\end{equation}

 \noindent the best-fit coefficients for our model can be found in Table \ref{tab:mod_coef}. The cubic boundary regression predicts values for \dmgal \ between 87.6 and 130.1\,\dmunits \ at $\sin|b|=0.67$ and $0.50$ respectively. We plot this model as the green dotted line on Figure \ref{fig:highlat}. When considering the error associated with this estimate of the structure of \dmgal \ across Galactic latitude, it is important to consider not only the error in estimation of fit parameters, but also the error introduced by the scatter in \dmhalo \ over Galactic longitudes. We provide pointwise bootstrap errors implied for the fit parameters\footnote{\url{https://www.canfar.net/citation/landing?doi=22.0079}}. 
 These boundary estimates are summarized for their comparison in Section \ref{sec:halomodels} to existing models of \dmhalo \ in Table \ref{tab:estimates}.

We plot \dmdisk \ in Figure \ref{fig:highlat} as estimated by the two popular free electron density models, NE2001 \citep{ne2001} and YMW16 \citep{ymw16}. We again removed from the data lines-of-sight with \anedit{emission measures (EMs)} above the 95$^{\text{th}}$ percentile of the sky. We remove lines of sight outside of CHIME's sky coverage as in Figure \ref{fig:nolines} to ensure an appropriate comparison. NE2001 and YMW16 are shown in blue and pink shaded regions on Figure \ref{fig:highlat} representing the range of maximum Galactic contributions over Galactic longitudes. The dashed lines of the same color located within the shaded regions of both models represent the median value over the relevant Galactic longitude. The implied DM of the WIM disk component as a function of Galactic latitude $b$ found by \cite{ocker}, shown in Equation \ref{eqn:ocker}.

\noindent is shown in yellow in Figure \ref{fig:highlat}, where the solid line represents the best-fit model and the surrounding region represents the model fit uncertainty. The estimated \dmdisk \ from YMW16, NE2001, and \cite{ocker} mostly bound the DMs of the pulsars and RRATs from above. There are a few exceptions where an observed Galactic source is only a few \dmunits \ above the largest expected \dmdisk \ from a given model at the relevant Galactic latitude. The FRB DMs are all $>30$\,\dmunits \ larger than the largest expected \dmdisk \ from any model. 

In summary, our constant Galactic contribution model, constant \dmhalo \ model, LOWESS Boundary estimate and Cubic Boundary estimate result in high-latitude upper limits on \dmgal \ from \dmmeightyone \ to 141\,\dmunits \ depending on the model and line of sight considered. By subtracting the \dmdisk \ estimate from \cite{ocker} assuming slab geometry of the disk, we can place upper limits on the \dmhalo \ alone. These constraints range from 52 to 111\,\dmunits \ depending on the Galactic latitude and model considered.

\begin{deluxetable*}{lll}
\label{tab:estimates}
\caption{Summary of the our boundary (upper limit) estimates of \dmhalo \ and \dmgal \ as presented in Section \ref{sec:analysis}. \dmhalo \ upper limits  are derived from the \dmgal \ estimates by subtracting an estimate of \dmdisk \ from \cite{ocker}, shown in Equation \ref{eqn:ocker}. }
\tablehead{
\colhead{Model} & \colhead{\dmgal \ upper limit range} & \dmhalo \ upper limit range \\
 & (\dmunits) & (\dmunits) }
\startdata
Constant \dmhalo & $76-99$ & 52 \\
Cubic boundary estimate & \dmmeightyone$ - 130$ & $52-83$ \\
LOWESS estimate & $88-141$ & $52-111$
\enddata
\end{deluxetable*}

\section{Discussion}
\label{sec:discussion}

\subsection{Potential Biases in Sample Collection}
\label{sec:biases}

We explore whether or not the gap in DM between disk pulsars and extragalactic FRBs is physical and caused by the Galactic halo. Note that each measured DM from an FRB source represents an upper limit on \dmhalo \ in that direction, regardless of the astrophysical significance of the lack of intermediate DMs. Our first analysis, which places upper limits across Galactic latitude, does not require the gap to be due to the presence of the halo in order to be a valid constraint.

We first discuss the potential biases contributing to this gap and then explore their effects. 

The first potential bias is that CHIME/FRB is less sensitive to radio bursts at low DMs. There are two effects contributing to the lower sensitivity. The first effect for this is, as mentioned in Section \ref{sec:data}, our pipeline only saves intensity data for radio pulses with DMs greater than \textit{at least one} of the maximal \dmgal \ estimates of YMW16 and NE2001 in their line of sight. However, as can be seen in Figure \ref{fig:highlat}, for high latitudes there is still a significant gap in radio pulse DM detections above the DM values where this condition would be relevant.

The second effect that makes CHIME/FRB less sensitive to low-DM bursts is that our wideband radio frequency interference (RFI) mitigation strategies preferentially remove signals from bright, low-DM events. This likely contributes to the apparent gap. We can quantify the extent of this and other system biases using studies of synthetic injected pulses (for more information on the injections system see \citealt{catalog} and \citealt{2022arXiv220614079M}). Using the injected pulse system we find that at excess DMs below 215\,\dmunits, the realtime pipeline recovers roughly 35\% of the injected pulses. This number only varies by 2\% between the region with DM excess less than 52\,\dmunits \ (where we have not detected FRBs) and the region with DM excess between 52 and 215\, \dmunits, with the pipeline recovering 2\% fewer events at the lower DM excess region. 

The second bias that could potentially explain this DM gap is a volume effect. \dmigm \ is believed to be the source of the observed Macquart (DM-$z$) relation, and hence should be a proxy for distance \citep{2020Natur.581..391M, 2021arXiv210108005J}.
In this way, modulo the variation coming from \dmgal \ and \dmhost, we expect to probe smaller volumes of space at lower DMs. 
However, if we restrict the DM range considered to smaller DMs, and hence volumes, there are fewer possible FRB hosts that could populate this DM region.

The last, and least-easily corrected, effect that could contribute to the apparent DM gap is the \dmhost \ of the FRBs. This is largely uncertain and estimates can easily vary from nearly zero (\citealt{m81rloc}) to hundreds of \dmunits \ (\citealt{2017ApJ...834L...7T}) depending on their location within and the properties of their host galaxies \citep[see][for more specific constraints]{2022ApJ...931...88C,2021arXiv211007418N,2022ApJ...927...35C}. The majority of FRBs do not have a known host galaxy. In addition, an estimate for \dmhost \  could include contributions from ionized gas local to the FRB source depending on the assumed-progenitor of the FRB. From the perspective of this analysis, \dmhost \ and \dmhalo \  contributions are degenerate. 
Without additional knowledge of their local environments, each of the considered FRBs' total measured DMs (which are less than 250 \dmunits) could be attributed entirely to a host like that of FRB 20121102A, for example, which has an estimated \dmhost $\lesssim 342$\,\dmunits \, \citep{2017ApJ...834L...7T} or, that of repeating FRB 20190520B, for which \cite{ocker_host} infer a \dmhost $= 1121^{+ 89}_{-138}$ \dmunits. 

\anedit{In the Appendix we estimate the astrophysical significance of the `gap' by quantifying the likelihood of observing zero events within the gap. The overall conclusion from the conservative probability estimate is that the observed gap in DM is roughly consistent with arising from pipeline biases and volume effects alone. As any extragalactic DM from an FRB represents an upper limit on \dmhalo \ in that direction, this does not invalidate the upper limits we have placed as a function of Galactic latitude. Instead, it suggests through February 2021, the high-latitude FRB DMs detected by CHIME/FRB are consistent (under the stated assumptions) with the Galactic halo contributing 0 \dmunits \ to the total DM of the FRBs. Of course, the upper limit analysis we present supports values from 0 to minimally 52 \dmunits \  and maximally 111 \dmunits \ (depending on the sightline and model selected), favoring no value within that range.}

\anedit{Given that \dmhalo\,=\,0\,\dmunits \ is supported in our models and the `gap' analysis, one could argue that our sample is not yet of adequate size or resolution to detect the halo's total mass and extent, but rather only constrains it.}

\subsection{Model uncertainties and unmodelled contributions} 
\label{sec:errors}
There are three main sources of error when describing the boundary of the halo as in Section \ref{sec:analysis}. There is random error in parameter estimation, error due to unmodelled longitudinal variation, and the contribution of \dmhost \ and \dmigm. 

\begin{figure*}
\centering
\includegraphics[width=0.91\textwidth]{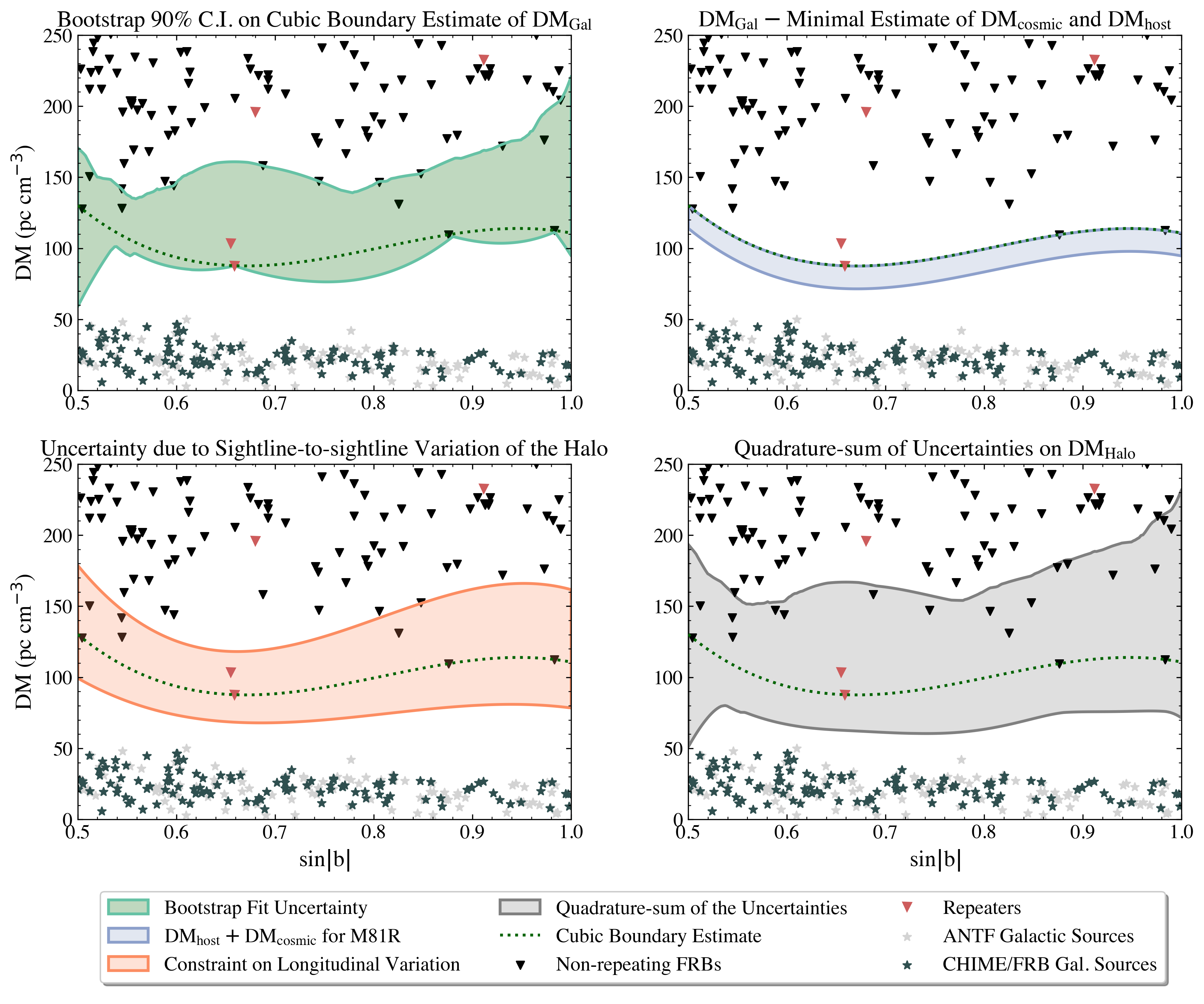}
\caption{\label{fig:errors} DM versus $\sin|b|$ of  radio pulse sources and one of the FRB-derived models of \dmgal \ from this work, to illustrate uncertainties in these models. We plot, in all panels, the cubic boundary estimate of \dmgal \ derived in Section \ref{sec:analysis} less the estimate of \dmdisk \ from \cite{ocker} (green dotted).
All panels are as in Figure \ref{fig:nolines}, but with upper-limit uncertainty regions demonstrated. These uncertainty regions show where the upper limits on \dmgal \ could lie, that is, where lines can be drawn such that all \dmgal \ smaller than that value would be supported by our data. \textit{Top Left:} The green region represents the 90\% confidence interval on the cubic boundary estimate via bootstrapping. That is, we remove one of our FRBs at random and re-estimate the cubic-polynomial boundary as described in Section \ref{sec:analysis}. This process is repeated 1000 times before the 90\% pointwise confidence intervals are selected and show in this panel.
\textit{Top Right:} The upper limits we report on \dmgal \ as a function of Galactic latitude assume that \dmhost \ and \dmigm \ are zero. For M81R we have a reasonable estimate of each of \dmgal, \dmhost, and \dmigm \ given its precise localization \citep{m81r, m81rloc}. We can look at the impact of this assumption using a minimal (conservative) estimate of \dmhost \, + \dmigm \, $=15$\,\dmunits \ value from M81R. We show the resulting \dmgal \ (blue region) after removing the implied minimal estimate of contamination by \dmhost \, + \dmigm \ from our cubic boundary estimate of \dmgal.  By removing this value of $15$\,\dmunits \ across the sky, it is assumed that each FRB in the sample has a greater \dmigm \ and \dmhost \ contribution than M81R, which may be appropriate given the exceptional nature of the M81R sightline (see Section \ref{sec:errors} for more details).
\textit{Bottom Left:} The region bounded by orange lines is a constraint on longitudinal variation of \dmhalo \ at each line of sight, given that \cite{2020ApJ...888..105Y} estimate the scatter of \dmhalo \ across the entire sky to be approximately 0.2 dex and that the variation in longitude must be a subset of the total sightline-to-sightline variation across the sky. Our four models for \dmgal \ are upper limits that do not account for \dmigm \ or \dmhost. This does not account for spherical geometry, that is, at very high latitudes we expect the sightline-to-sightline variation to become negligible as the sky area is also decreasing below spatial scales where we expect the halo plasma to vary. 
\textit{Bottom Right:} The three sources of uncertainty from the other panels added in quadrature. These three sources of uncertainty are not independent (for example, the some of the uncertainty in the cubic boundary estimate certainly arises from the sightline-to-sightline variation), so this error is larger and hence more conservative than the true combined error.}
\end{figure*}

As discussed in Section \ref{sec:analysis} when introducing the cubic boundary estimate of \dmgal, the first source of uncertainty is due to parameter estimation. We resample our original dataset of pairs of FRB DMs and latitudes, that were used to fit our \dmhalo \, models, 1000 times with replacement (bootstrapping) to estimate pointwise 90\% confidence intervals. We show this 90\% confidence interval of the cubic boundary estimate of \dmgal\ as the region bound by solid green lines in Figure \ref{fig:errors}.

The second source of uncertainty, also discussed in Section \ref{sec:analysis}, is longitudinal variation in \dmhalo. This variation introduces error in the models, which is unaccounted for (due to small sample size) in this analysis.

A scatter of 0.3$-$0.4 dex is seen in both \textit{Suzaku} \citep{2018ApJSuzaku} and HaloSat X-ray EM data \citep{Kaaret2020Nat} of the MW halo. If we knew exactly what fraction of this scatter can be attributed to the fluctuation of the MW halo gas density, it would tell us the approximate scatter of \dmhalo, as EM is proportional to the path integral of $n_e^2$ while DM is proportional to the path integral of $n_e$. \anedit{It is worth noting that that instrumental limitations of X-ray telescopes are such that these observations are sensitive to only the densest hot gas, whereas the integrated \dmhalo \ will include more distant, diffuse gas \citep{2006ApJ...644..174F,2007ApJ...658.1088Y}. As such, one may expect that \dmhalo \ have much less scatter.}  We assume that the an upper limit on the fluctuation of \dmhalo \ is approximately $\approx$ 0.2 dex.
This also constrains the total amount of longitudinal variation we expect. For each $\sin|b|$ we illustrate in Figure \ref{fig:errors} the extent of a 0.2 dex variation around our cubic boundary estimate of \dmgal\ (region bounded by solid orange lines; see Section \ref{sec:results} for more information on the cubic boundary estimation). 



To investigate the third source of error, that due to each FRB's non-zero and unaccounted for \dmigm \ and \dmhost, one can study the most constraining FRB sight-line, that of M81R. M81R is exceptional both in being the lowest-DM source in our sample, and being precisely localized within a globular cluster on the outskirts of the halo of M81 \citep{m81r, m81rloc}. In \cite{m81r}, the authors discuss the uncertainties in estimating this source's exact \dmhost \ and \dmigm, but ultimately conclude a minimal, conservative, expected \dmhost + \dmigm$ = 15 $\,\dmunits. 

Even given the conservative nature of the quadrature sum of these three sources of uncertainty, the region does not encompass any of the Galactic sources in our sample. This is indicative of the conservative nature of the upper limits presented in the paper, and is a nice consistency check for our models.  

Given that the source is relatively nearby (so we expect very little \dmigm) and has essentially no local or disk DM contributing to \dmhost, it could be argued that this is an edge case of the FRB population and it would be appropriate to subtract this same lower bound on \dmigm + \dmhost$=15$ \dmunits \ from every line of sight. We refrain from making this generalization in our \dmhalo \ estimates given our sample is not universally localized to the precision required to estimate the \dmhost \ contribution, nor is there sufficient knowledge about the population of FRB hosts to make a meaningful distribution-based argument. We still demonstrate, however, the magnitude of this minimal expected contamination of \dmhost + \dmigm \ in Figure \ref{fig:errors} (region bounded by a solid blue line).

\subsection{Constraints on Existing Halo Models}
\label{sec:halomodels}

Our study was motivated by wanting to observationally constrain the gas content of the MW halo to distinguish between different galaxy formation models. 
We obtain this constraint by comparing the upper limits implied by FRBs to the estimates made by models with different physical assumptions. First we review these models and compare their estimates to our \dmhalo \ boundary estimates. Table \ref{tab:estimates} summarizes our \dmhalo \ boundary (upper limit) estimates from Section \ref{sec:analysis}.

\cite{keatingpen} compute most of these estimates of \dmhalo \ using gas profiles of halo models. Two total masses of the MW halo are considered in each of their estimates, \halomass $= 1.5 \times 10^{12} M_\odot$ and \halomass $= 3.5 \times 10^{12} M_\odot$. We compare our FRB constraints with the range bounded by the halo model estimates from the lower and higher mass scenario\footnote{the fraction of ionized gas differs and is specified by each model}. In addition, for some of the models multiple physical scenarios are considered. In this case, which is noted in the brief descriptions of the models that follow, we additionally consider the range in values between these multiple scenarios. We briefly summarize the models included for their comparison to our upper limits. 

\begin{figure*}
    \centering
\includegraphics[width=\textwidth]{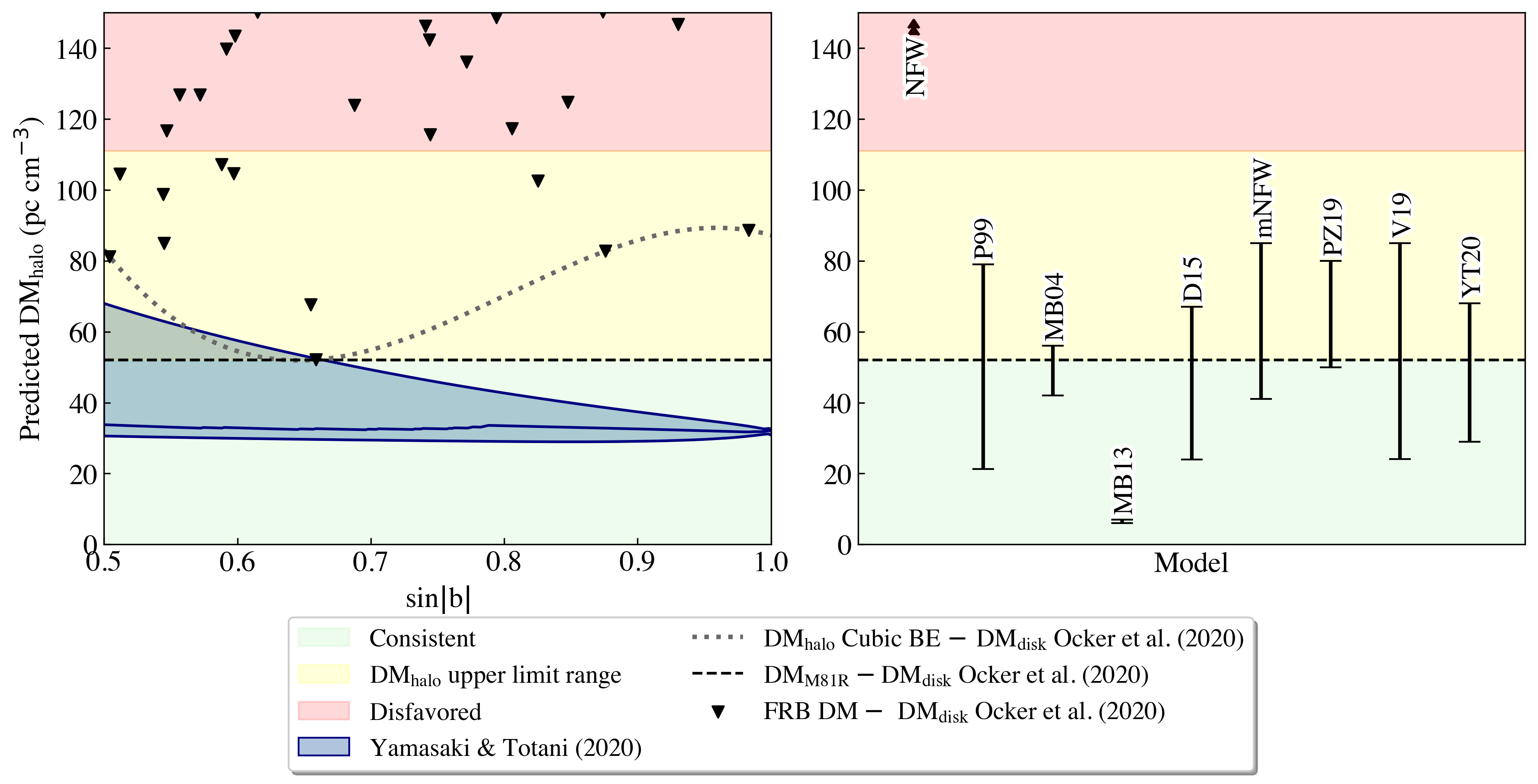}
    \caption{
    \textit{Left panel:} Comparisons between the predicted \dmhalo \ vs Galactic latitude of the upper limits derived from FRBs in this work and \cite{2020ApJ...888..105Y}.
   The red region shows values of \dmhalo \ that are contradictory to our observations as they are higher than our predictions for \dmhalo \ at all latitudes. The yellow region shows the span of our \dmhalo \ model predictions ($52-111$\,\dmunits) and hence denotes \dmhost \  values where models are not strictly ruled out but seem less likely than models in the green region due to the conservative upper-limit nature of our result. 
   The dashed line shows the DM of the M81 repeater \citep{m81r} minus the mean prediction for \dmdisk \ at the source's latitude from \cite{ocker} assuming a slab geometry. Black triangles represent the Galactic latitude and excess DM of FRBs in our dataset, where excess DM is defined here as the true DM minus \dmdisk \ as estimated by \cite{ocker} assuming a slab geometry (Equation \ref{eqn:ocker}). We also plot the cubic boundary estimate of \dmgal \ derived in Section \ref{sec:analysis} less the estimate of \dmdisk \ from \cite{ocker} (gray dotted). YT20 make predictions for the halo contribution as a function of Galactic latitude and we show these predictions in the blue region, where the span represents the range of predictions over all Galactic longitudes in CHIME/FRB's field of view. The three solid blue lines indicate the minimum, median, and maximum \dmhalo \ at each $\sin|b|$ for all considered $l$. At $l,b  = 142.19^\circ, +41.22^\circ$, the position of M81R, the \dmhalo \ prediction from YT20 is 30.6 \dmunits. 
    \textit{Right panel}: Comparisons between the predicted \dmhalo \ for a selection of popular halo models (ordered by publication date) and the upper limits derived from FRBs in this work. The acronyms for the models are defined along with their brief descriptions in section \ref{sec:halomodels}. For MB13, V19, MB04, and NFW models the ranges represent the different input masses for the Milky Way halo spanning $(1.5-3.5) \times 10^{12} M_\odot$ \citep{keatingpen}. The range for P99 includes three values of heated core radius. The range in YT20 represents the longitudinal variation in the high latitude portion of the model.}
    \label{fig:halo}
\end{figure*}

\paragraph{\cite{nfw} and \cite{2017ApJ...846L..24M} (mNFW)}
The Navarro-Frenk-White (NFW; \citealt{nfw}) profile describes well the density profile of virialized dark matter halos in cosmological simulations. A simple model for the baryonic matter is to assume that it traces dark matter near the cosmic ratio \citep[$\Omega_b/\Omega_m \sim 0.2 $;][]{2016A&A...594A..13P} down to ten percent of the MW virial radius, in which case the gas density profile ($\rho$) as a function of distance from the center of the Galaxy ($r$) can be described using the NFW profile, 

\begin{equation}
    \rho(r) = \frac{\rho_0}{y(1+y^2)} 
\end{equation}
where $y = c(r/r_{V})$ with concentration $c$ and $r_{V}$ is the virial radius, and $\rho_0$ is a characteristic density. This model predicts \dmhalo $\approx 300-500$ \dmunits \ and hence was inconsistent with previous observations \citep{keatingpen}, and remains inconsistent given our observations. This simple model does not account for non-linear effects facilitated by, e.g., feedback, accretion, and shocks. \cite{2017ApJ...846L..24M} modify the NFW profile with an additional two parameters ($y_0, a$) based on measurements of \ion{O}{6} absorption in quasar spectra caused by intervening galactic halos: 

\begin{equation}
    \rho(r) = \frac{\rho_0 }{y(y_0+y)^{2+\alpha}}. 
\end{equation}
This extension to baryonic matter accounts for feedback.
We consider (as in \citealt{prochaskaandzhang} and \citealt{keatingpen}) profiles with $y_0=2$ and $y_0=4$ in the span of this modified NFW profile (mNFW) predicted \dmhalo \ in Figure \ref{fig:halo} while keeping $\alpha = 2$ fixed for both cases. In the $y_0=2$ case, the profile is disfavored for both the halo masses considered (i.e., between $1.5 - 3.0 \times 10^{12} M_\odot$) as it predicts \dmhalo $= 66-86$\,\dmunits, which is higher than the upper limits in most lines-of-sight of each of our four models. The profile with $y_0=4$ remains more plausible for lower masses, as it predicts \dmhalo $= 41-51$\,\dmunits.

\paragraph{\cite{2004MNRAS.355..694M}} MB04 create their gas density profile by assuming that the halo gas is adiabatic and in hydrostatic equilibrium, taking into account the expectation that the hot gas in halos is prone to fragmentation during cooling due to its thermal instability. The resulting density profile is defined as 

\begin{equation}
    \rho(r) = \rho_c \left( 1 + \frac{3.7}{y} \ln(1+y) - \frac{3.7}{C_c} \ln(1+C_c) \right)^{3/2}
\end{equation}
where again $y = c(r/r_{V})$. $\rho_c$ is a normalization constant set by the assumed gas mass of the halo, and  $C_c = c \frac{r_c}{r_{V}}$ with $r_c = 147$ kpc as assumed by \cite{keatingpen} and \cite{prochaskaandzhang}. All masses considered are compatible as they are slightly lower than the upper limits given by our observations, as \dmhalo \ is estimated to be 42, 56 \dmunits \ in the low and high halo mass scenarios respectively. If this model is correct and the mass of the MW halo is within $(1.5, 3.5) \times 10^{12} M_\odot$, when this estimate is used in the line of sight of M81R it would suggest that \dmhost \ (including that from the likely significant fraction of M81's halo which the burst encounters) would need to contribute considerably less DM than the MW halo.

\paragraph{\cite{MB13}}
 MB13 use archival soft X-ray data from \textit{XMM-Newton}'s Reflection Grating Spectrometer to measure \ion{O}{7} K$\alpha$. This is used to find best fit parameters ($n_0 = 0.46$ cm$^{-3}, r_c = 0.35 \text{ kpc and } \beta =  0.71$) for an underlying spherical density of the hot Galactic halo ($n$) model of the form 

\begin{equation}
    n(r) = n_0\left( 1 +\frac{r}{r_c}  \right)^{-3 \beta/2}
\end{equation}
with the addition of an ambient density component due to ram-pressure stripping of $n = 1 \times 10^{-5}$ cm$^{-3}$ out to 200 kpc. As the density profile with the lowest estimated \dmhalo, this model is not ruled out at these masses. \cite{keatingpen} estimate the contribution from these $\approx 6$ \dmunits \ in the low mass halo scenario and $\approx 7$ \dmunits \ in the higher mass scenario.

\paragraph{\cite{P99}}

 P99 uses an entropy-floor singular isothermal sphere model motivated by observations of the soft X-ray background. In this model, the halo gas is assumed to have two phases: an outer region in which gas traces mass isothermally; and an inner region in which the gas has been heated to constant entropy, invoking baryonic feedback. \cite{keatingpen} consider two cases of the model, one with a heated core radius ($r_c$) that produces X-ray emission at the limit of the observational constraints of \cite{2003ApJ...588..696M}, 
 and one which maximizes the effect of feedback by choosing $r_c$ equal to the virial radius of the MW. We consider each of these profiles in Figure \ref{fig:halo}. When \halomass $= 1.5 \times 10^{12} M_\odot$ is assumed, and $r_c = 0.34 \, r_{V}$ in order to match the X-ray emission, \cite{keatingpen} estimate \dmhalo $\approx 79$ \dmunits \ which is larger than some of our upper limits and hence is largely inconsistent with our observations. When \halomass $= 3.5 \times 10^{12} M_\odot$ and $r_c = 0.86 \, r_{V}$ (predicted \dmhalo $= 34$\,\dmunits) the measurements are consistent with our observations. Similarly, in either the high (\halomass $ = 3.5 \times 10^{12} M_\odot$) or low ($= 1.5 \times 10^{12} M_\odot$) mass scenario, when the heated core radius $r_c$ is set equal to $r_{V}$ (\dmhalo $=28,  21$\,\dmunits \ for the high, low mass scenario respectively) the results are consistent with our observations.

\paragraph{\cite{V19}}
 V19 constructed a model for the halo, called the pNFW model, which assumes a confining gravitational potential with a constant circular velocity at small radii. At larger radii the circular velocity profile is assumed to decline like that of a NFW halo with scale radius $r_{s, \text{ NFW}}$. These two profiles are joined continuously at a radius of 2.163\,$r_{s, \text{ NFW}}$. The author provides a table of formula coefficients as a function of input halo mass for the resultant density profile. In this model, only the lower halo mass scenarios are consistent with all of our lines of sight since the model estimates \dmhalo$=24-84$\,\dmunits \ for MW halo masses between $1.5-3.0 \times 10^{12} M_\odot$.\\

In addition to the estimates derived by \cite{keatingpen} from the above density profiles, we compare our observations to the following estimates for \dmhalo.

\paragraph{\cite{2020ApJ...888..105Y}}

 YT20 model the MW halo with a spherical component of isothermal gas in hydrostatic equilibrium and a disk-like hot gas component to reproduce the directional dependence of X-ray emission measure observed by \cite{2018ApJ...862...34N}. They present an analytic formula for \dmhalo \ that we plot as a function of Galactic latitude while representing the longitudinal variation as a span in the left panel of Figure \ref{fig:halo}. At $|b| > 30^{\circ}$, this model predicts \dmhalo \ of between $29$\,\dmunits \ and 68\,\dmunits \ depending on the line of sight considered. As can be seen in Figure \ref{fig:halo}, at each $b$ the minimum and median value lie below our cubic boundary estimate of \dmhalo, and in most cases the maximum \dmhalo \ prediction also lies below our model. At the sky position of M81R, YT20 predicts \dmhalo $\approx 30.5$ \dmunits. Our constraint is \dmhalo $< 52$ \dmunits \ for all boundary models and hence we find the YT20 model is consistent with our FRB observations. 

\paragraph{\cite{dolag}}

D15 perform cosmological simulations of a MW-like galactic halo including hot thermal electrons in order to estimate \dmhalo. Their probable values for DM$_{\text{halo}}$, depending on which inner radius one expects from the edge of the Galactic disk, range over $\approx 24-67 $ pc\,cm$^{-3}$. This range, particularly for the larger radii of the edge of the Galactic disk remains highly relevant and agrees well with our observations, as does the commonly cited representative halo electron column estimate of \dmhalo $ = 30$ \dmunits \ selected by the authors. This representative value assumes integration radii beginning 17 kpc away from the Galactic Center, the maximal extent of NE2001 that was used by \cite{dolag} to model \dmdisk.

\paragraph{\cite{prochaskaandzhang}}

 PZ19 look at tracers of the `hot' ($T \sim 10^6K$) and `cool' ($T \sim 10^4$K) components of the halo gas. These tracers, namely observations of \ion{O}{6} and \ion{O}{7} absorption \citep{2015ApJS..217...21F}, Si $\textrm{II}$ and Si $\textrm{III}$  \citep{si23}, and high velocity clouds \citep{hi4pi} are combined with hydrostatic models of the halo to estimate \dmhalo\ $ = 50 - 80$ pc\,cm$^{-3}$ integrated to 200 kpc. This is within the upper limit range of our various models, but most of the range is above the excess DM of M81R (see also \citealt{m81r}).\\
 We compare our \dmhalo \ boundary estimates and upper limits to estimates of \dmhalo \ implied by various models in Figure \ref{fig:halo}.  The red region in Figure \ref{fig:halo} shows the DM range that cannot be supported by our observations regardless of model chosen or line of sight. Within the yellow region, we show the DM range that encompasses all upper limits from our models, ranging between 52 and 111 \dmunits. We highlight the excess DM of M81R (FRB 20200120E; \citealt{m81r}), which is the lowest extragalactic DM in our sample, with the black dotted line.

To summarize, the NFW profile can be unambiguously ruled out (as was previously known, e.g.,  \citealt{2013ApJ...762...20F} and \citealt{P99}), and each of MB13, YT20, and D15 are consistent with our observations. In the case of mNFW, MB04, and V19, the models are mildly in tension with our observations for the high MW halo mass considered ($3.5 \times 10^{12} M_{\odot}$), but not for the low mass ($1.5 \times 10^{12} M_{\odot}$) scenario. Similarly, for P99, the model is supported only when the heated core radius is set at the virial radius or the MW halo is assumed to be lower mass. The majority of the \dmhalo \ range proposed by PZ19 is higher than our estimates, but remains  possible in the scenario there is significant \dmhalo \ scatter in the sky, as acknowledged for the M81R sightline \citep{m81r}.

 Baryonic feedback processes and their overall effect are still relatively uncertain in galaxy formation, however it is interesting to note that both the NFW/mNFW model and P99's models flip from inconsistent with our observations to consistent when consideration for the effect of feedback is increased. The cosmological simulation of D15 which results in a \dmhalo \ estimate that is in great agreement with our observations, also accounts for the energy released in explosions of massive stars as supernovae, a type of baryonic feedback, and for feedback from active galactic nuclei.

\section{Conclusions}
\label{sec:conclusions}
We explore the constraints on the total Milky Way (MW) dispersion measure (DM) as well as the MW halo DM using CHIME/FRB's large, extragalactic, fast radio burst (FRB) source population. This sample of DM measurements offers a unique opportunity to constrain the distribution of the Galactic plasma and estimate the MW halo DM contribution upper limits as a function of Galactic latitude.
The observation-based high-latitude upper limits on the Galactic DM contribution range over \dmmeightyone--141 \dmunits \ depending on the chosen model and the Galactic latitude of interest. Subtracting estimates of the disk contribution from \cite{ocker}, we derive upper limits on the MW halo DM contribution ranging over 52--111 \dmunits. \anedit{These results agree with the recently reported constraint of \dmhalo$\leq 47.3$\,\dmunits\, along the line-of-sight toward FRB 20220319D, located at a comparatively low Galactic latitude of $b\sim +9.1^{\circ}$ \citep{2023arXiv230101000R}.}

Although there is a DM gap between Galactic and extragalactic radio pulses, assuming the rate at which FRB sources are detected can be described using Poisson statistics, and using measured population statistics from the first CHIME/FRB catalog, we find that this lack of intermediate DM radio sources is compatible with having arisen from volume effects and pipeline bias alone. The presence of the gap is therefore not evidence of a \dmhalo \ contribution that is non-zero.

Our constraints on the MW halo DM contribution seem at tension with most popular estimates of \dmhalo \ \citep[e.g.,][]{2004MNRAS.355..694M, 2017ApJ...846L..24M, V19} when assuming a MW halo mass of $3.5 \times 10^{12} M_{\odot}$, with the exception of \cite{MB13} and \cite{P99}. In part, this tension arises as our estimates are necessarily an overestimate of the true value, as we do not estimate and remove DM contributions from the intergalactic medium nor the host galaxy of each FRB. If we assume a lower MW halo mass estimate of of $1.5 \times 10^{12} M_{\odot}$, our constraints agree with more models, including those proposed by \cite{2004MNRAS.355..694M, 2017ApJ...846L..24M} and \cite{V19}. The estimates of the MW halo DM contribution produced by \cite{dolag} using cosmological simulations of a MW-like galactic halo are supported by our observations. So too is the MW halo model of \cite{2020ApJ...888..105Y}, which combines a spherical isothermal gas component and a disk-like component hot gas component.  The majority of the \dmhalo \ range proposed by PZ19 is higher than our estimates, but remains possible in the scenario there is significant \dmhalo \ scatter in the sky, as acknowledged for the M81R sightline \citep{m81r}. For some models these results seem to emphasize the importance of the role of baryonic feedback in Galaxy formation. 


 \anedit{While many models of the density of the halo gas invoke strict or quasi-spherical symmetry, one expects the ionized gas in the Local Group to be ellipsoidal, extended from our Galaxy towards M31 due to the inflows, outflows, and tidal interactions between our Galaxy and M31 \citep{2007ApJ...669..990B}. Simulation work \citep{2014MNRAS.441.2593N} also finds evidence for this gas excess between a pair of galaxies resembling M31 and the Milky Way compared to another, random, line of sight. In searches for an excess in DM from FRBs which intersect the dark matter halos of other galaxies, \cite{2021arXiv210713692C} find a higher excess DM in these lines of sight than expected from diffuse gas surrounding isolated galaxies. The authors suggest this DM excess is potentially due to ionized media in galaxy groups, including the Local Group. \cite{2022arXiv220904455W} present a similar analysis, but introduce a weighted-stacking scheme which minimizes the effect of the variance of the observed DM distribution and derive a significance for the result that is lower than that found by \cite{2021arXiv210713692C} (probability $>0.99$ vs. $>0.68$ to $>0.95$). We plan to repeat our study observed in two dimensions (i.e., producing a sky map) once the known FRB population has roughly doubled. This 2D map will allow us to search for evidence of asymmetries in the Galaxy or such an ellipsoidal halo gas distribution that is extended by interactions with our Galaxy group, expected to be dominated by interactions with M31. }


\section*{Acknowledgements}
We thank \anedit{Jo Bovy}, Stanislav Volgushev, and Jeremy Webb for discussions vital to preparing this work. \anedit{We are also grateful to the referee for their very thoughtful and constructive comments}. \\
A.M.C is funded by an NSERC Doctoral Postgraduate Scholarship. 
M.B. is a McWilliams Fellow.
The Dunlap Institute is funded through an endowment established by the David Dunlap family and the University of Toronto. B.M.G. acknowledges the support of the Natural Sciences and Engineering Research Council of Canada (NSERC) through grant RGPIN-2022-03163, and of the Canada Research Chairs program.
GME acknowledges funding from NSERC through Discovery Grant RGPIN-2020-04554 and from UofT through the Connaught New Researcher Award.
V.M.K. holds the Lorne Trottier Chair in Astrophysics \& Cosmology, a Distinguished James McGill Professorship, and receives support from an NSERC Discovery grant (RGPIN 228738-13), from an R. Howard Webster Foundation Fellowship from CIFAR, and from the FRQNT CRAQ.
K.W.M. holds the Adam J. Burgasser Chair in Astrophysics and is supported by an NSF Grant (2008031).
A.P.C is a Vanier Canada Graduate Scholar.
F.A.D is supported by the UBC Four Year Fellowship.
\anedit{C.L. was supported by the U.S. Department of Defense (DoD) through the National Defense Science \& Engineering Graduate Fellowship (NDSEG) Program.}
A.P. is funded by an Ontario Graduate Scholarship.
A.B.P. is a McGill Space Institute (MSI) Fellow and a Fonds de Recherche du Quebec -- Nature et Technologies (FRQNT) postdoctoral fellow.
Z.P. is a Dunlap Fellow.
The National Radio Astronomy Observatory is a facility of the National Science Foundation operated under cooperative agreement by Associated Universities, Inc. S.M.R. is a CIFAR Fellow and is supported by the NSF Physics Frontiers Center awards 1430284 and 2020265.
K.S. is supported by the NSF Graduate Research Fellowship Program.
FRB research at UBC is supported by an NSERC Discovery Grant and by the Canadian Institute for Advanced Research.
D.C.S. acknowledges the support of the Natural Sciences and Engineering Research Council of Canada (NSERC), RGPIN-2021-03985
We acknowledge that CHIME is located on the traditional, ancestral, and unceded territory of the Syilx/Okanagan people. We are grateful to the staff of the Dominion Radio Astrophysical Observatory, which is operated by the National Research Council of Canada.  CHIME is funded by a grant from the Canada Foundation for Innovation (CFI) 2012 Leading Edge Fund (Project 31170) and by contributions from the provinces of British Columbia, Qu\'{e}bec and Ontario. The CHIME/FRB Project is funded by a grant from the CFI 2015 Innovation Fund (Project 33213) and by contributions from the provinces of British Columbia and Qu\'{e}bec, and by the Dunlap Institute for Astronomy and Astrophysics at the University of Toronto. Additional support was provided by the Canadian Institute for Advanced Research (CIFAR), McGill University and the McGill Space Institute thanks to the Trottier Family Foundation, and the University of British Columbia.

\appendix
\section*{\anedit{Quantitative analysis of the DM Gap}}
\label{app:prob}
Given the biases discussed in Section \ref{sec:biases} in this \anedit{Appendix} we will answer the question \textit{`Is this gap astrophysical?'}, or, equivalently, \textit{`Does one need more than volume and selection effects to explain this gap?'}. We do not assert what fraction of the gap can be attributed to \dmhalo \ and \dmhost \ accordingly. To derive the astrophysical significance, we will quantify the likelihood of observing zero events within the `gap', given the observation of 93 FRB sources in the remaining DM range of our sample and considering only the volume and pipeline biases. In order to estimate this likelihood, we make some simplifying assumptions, and highlight these assumptions as they appear in the derivation. We show at the end of the section that ultimately these assumptions result in a conservative estimate. 

First we define DM$-$\dmdisk \ as the excess DM, and assume it is contributed solely by the IGM. That is, we are assuming there is no contribution from the MW halo (\dmhalo\,$=0$ \dmunits) and no contribution from the host galaxy (\dmhost \,$= 0$ \dmunits). If we assume both are zero and we know that in a given line of sight the \dmigm \ is proportional to distance ($d$) and distance is proportional to redshift ($z$)\footnote{This can only be assumed for $z\ll 1$, where space is approximately Euclidean.}, we are essentially extending the volume in which FRB sources can exist right to the edge of our MW WIM disk. We can estimate the relative rate of FRBs between two volumes using the fluence distribution (commonly referred to as $\log(N) / \log(F)$) of FRBs. 
We will compare the relative rate of FRBs in the DM gap (excess DMs $[0, 52)$ \dmunits) and in the rest of the sample, which spans excess DMs from $[52, 215]$ \dmunits. 

To simplify, assume FRBs are standard candles, that is, each burst has equal intrinsic energy.
The number of FRBs ($N$) contained in a given spherical volume of radius $d$ is 

\begin{equation}
\label{eqn:powerlaw}
    N \propto F^\alpha \propto \left(d^{-2}\right)^{\alpha}
\end{equation}
where $F$ is the FRB fluence, $d$ is distance, and $\alpha$ is power-law index for the cumulative fluence distribution $(\alpha < 0)$. For a non-evolving population in Euclidean space, one expects $\alpha=-3/2 $, and this is in agreement with the $\alpha = - 1.40\pm0.11(\text{stat.})^{+0.06}_{-0.09}(\text{sys.})$ measured by CHIME/FRB when including bursts at all DMs/distances in the first FRB catalog \citep{catalog}. At small $d$, where space is approximately Euclidean, we can assume $d \propto z$ and hence 

\begin{equation}
    N(<z) \propto z^{-2\alpha}.
\end{equation}

Now compare the ratio of the volume in which we detect no FRBs (the gap, v$_1$) and the volume containing our FRB sample (v$_2$). We can estimate the redshifts at the DM excesses which define our volumes of interest (52 and 215 \dmunits) as in \cite{2020Natur.581..391M}, who assume cosmological parameters as measured by  \cite{2016A&A...594A..13P}. The expected relation between \dmigm \, and redshift results in redshift estimates of $0.06$ at $52$ \dmunits, and $0.23$ at 215 \dmunits. These redshift estimates have uncertainty due to scatter within the IGM on the order of the estimates ($0.04, 0.11$ for the first and second volume boundary, respectively) according to the 90\% confidence interval on the fit of the \dmigm--$z$. However, we simply select the central value of the redshift estimate to define our boundaries and discuss the effect of the scatter in the IGM on this calculation at the end of this section. 


We expect the ratio of number of sources detected in the first and second volumes to be 

\begin{align}
\frac{N_{\text{FRBs, v$_1$}}}{N_{\text{FRBs, v$_2$}}} & =
\frac{N_{\text{FRBs}}\left(z \in [0, 0.06)\right)}{N_{\text{FRBs}} \left(z \in [0,0.23] - N \in [0, 0.06)\right)} \\
    &= \frac{z_1^{-2\alpha}}{z_2^{-2\alpha} - z_1^{-2\alpha}} \\ 
    & = \frac{1}{\left(\frac{z_2}{z_1}\right)^{-2\alpha} -1 } \label{eqn:expectedfrac},
\end{align}

where $N_{\text{FRBs}}$ describes the number of detectable FRBs in a given volume or redshift range, and $z_1 = 0.06$ and $z_2 = 0.23$ are the redshifts that define boundaries of the two volumes of interest (v$_1$, v$_2$ respectively). 

Finally we must account for the sensitivity of CHIME/FRB to radio pulses in the DM range considered. We correct for this effect using information from CHIME/FRB's synthetic signal injection system. For injected FRB signals with excess DM $\in [0,52)$ and $\in[52,250]$, the fraction of detected events ($\mu$) are $\mu_{\text{v$_1$}}=$ 0.346 and $\mu_{\text{v$_2$}}=$ 0.366 respectively. Hence we adjust Equation \ref{eqn:expectedfrac} to account for this bias

\begin{align}
    \frac{N_{\text{det, v$_1$}}}{N_{\text{det, v$_2$}}}  & =  \frac{\mu_{\text{v$_1$}} N_{\text{FRBs v$_1$}}}{\mu_{\text{v$_2$}}N_{\text{FRBs v$_2$}}} \\
    & = \frac{\mu_{\text{v$_1$}}}{\mu_{\text{v$_2$}}} \frac{1}{\left(\frac{z_2}{z_1}\right)^{-2\alpha} -1 },
\end{align}

where $N_{\text{det}}$ is the number of FRBs we expect CHIME/FRBs realtime pipeline to detect. We only use the FRBs that were detected while the CHIME/FRB pipeline was in the configuration these injections are intended to gauge, leaving us with 83 of our sample 93 FRBs. The other 10 FRBs were detected after November 2020 when a significant change was implemented in our realtime dedispersion alogrithm.

Since we have measured a non-zero $N_{\text{det, v$_2$}}$, we estimate the expected $N_{\text{det, v$_1$}}$ in the same time frame. Treated as a rate, we can then use Poisson statistics to describe the likelihood of our observation of zero events in volume 1 under our initial assumptions,

\begin{align}
   N_{\text{det, v$_1$}} & = \frac{\mu_{\text{v$_1$}}}{\mu_{\text{v$_2$}}} \frac{N_{\text{det, v$_2$}}}{\left(\frac{z_2}{z_1}\right)^{-2\alpha} -1 }.
\end{align}

The probability of detecting no events given a rate of $ N_{\text{det, v$_1$}}$ assuming FRB source detection can be modelled as a Poisson process, is 
\begin{align*}
    P(\text{no events}\,| \text{ rate} =  N_{\text{det, v$_1$}}) & = e^{- N_{\text{det, v$_1$}} }
\end{align*}
where $P$ is the likelihood of the observation.

The probability of obtaining 0 events in volume 1 under the null hypothesis of there being no astrophysical DM gap is therefore:
\begin{align}
    P & = \text{exp} \left[ -  \frac{\mu_{\text{v$_1$}}}{\mu_{\text{v$_2$}}}\frac{N_{\text{det, v$_2$}}}{\left(\frac{z_2}{z_1}\right)^{-2\alpha} -1 }\right] \label{eqn:prob} \\
    & =  0.24. 
\end{align}

Hence, under these assumptions the resulting likelihood suggests that the lack of FRB detections in the first volume, considering the number of FRBs detected in the second volume, is consistent with being due to pipeline biases and volume effects alone. 

If we look at the assumptions that we have made, we find that this likelihood is quite conservative (i.e., likely an overestimate, hence we expect to see zero FRBs in this region due to volume and selection effects alone less frequently than 24\% of the time if we could repeat the experiment again and again). 
The assumption that FRBs are standard candles predicts fewer low-fluence bursts compared to the true underlying luminosity distribution as seen in the first CHIME/FRB catalog (\citealt{catalog}) as well as in observations of repeaters \citep[e.g.,][]{2022ApJ...927...59L,2021Natur.598..267L}. As volume 1 is closer than volume 2, the omission of these low-fluence bursts would decrease the number of detectable bursts per unit volume more in volume 1 than volume 2. That is, if there were a population of intermediate fluence FRBs, there should be bursts that are detectable in volume 1 and not volume 2. Hence, we can safely assume the fraction of detected bursts in volume 1 compared to volume 2, and the true rate, would be larger than what we estimate. This underestimation of the rate will overestimate the likelihood of our observation, and hence is a conservative estimate. 

In \cite{catalog}, the authors investigate the DM--distance relation by splitting the FRB sample into ``low-DM'' ($100-500$\,\dmunits) \ and ``high-DM'' (above 500\,\dmunits) and measuring the $\alpha$ for the subsets. One might believe, then, a more appropriate $\alpha$ to be used would be that measured by \cite{catalog} who infer $\alpha = -0.95\pm0.15(\text{stat})^{+0.06}_{-0.19}(\text{sys})$ for events with DMs $100- 500$ \dmunits.  However, since we have assumed a standard candle luminosity function, it is not appropriate to use this $\alpha$ value.

Additionally, when we estimate the redshifts that correspond to \dmigm \ $=52, \,  215$\,\dmunits \ respectively, hence defining our volume boundaries, we do not consider the uncertainty presented by \cite{2020Natur.581..391M}. These uncertainties represent the 90\% confidence intervals on the fit of \dmigm\, vs redshift (DM$-z$), accounting for scatter within the IGM (cosmic structure). However, as above, this variance is more likely to result in additional FRBs being detected in the first volume, so our estimate remains conservative.

\bibliography{sample63}{}
\bibliographystyle{aasjournal}
\end{document}